%% file: template.tex
\documentclass{article}

\usepackage{arxiv}

\usepackage[utf8]{inputenc} % allow utf-8 input
\usepackage[T1]{fontenc}    % use 8-bit T1 fonts
\usepackage[
    colorlinks=false,
    linkcolor=blue,
    hidelinks,
    breaklinks]{hyperref}       % hyperlinks
\usepackage{url}            % simple URL typesetting
\usepackage{booktabs}       % professional-quality tables
\usepackage{amsfonts, amsmath}       % blackboard math symbols
\usepackage{nicefrac}       % compact symbols for 1/2, etc.
\usepackage{microtype}      % microtypography
\usepackage{cleveref}       % smart cross-referencing
\usepackage{graphicx}
\usepackage{doi}

\title{Millisecond-scale volatile memory in HZO ferroelectric capacitors for bio-inspired temporal computing}

\newif\ifuniqueAffiliation
\ifuniqueAffiliation % Standard variant of author block
\author{ 
    \href{https://orcid.org/0000-0003-0993-5593}{\includegraphics[scale=0.06]{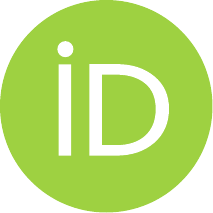}\hspace{1mm}Luca Fehlings}\thanks{Now with Zernike Institute for Advanced Materials \& Groningen Cognitive Systems and Materials Center (CogniGron), University of Groningen} \\
	\\
	NaMLab gGmbH\\
	01187 Dresden, Germany \\
	\texttt{Corresponding author: l.d.fehlings@rug.nl} \\
	\And
	\href{https://orcid.org/0000-0003-3814-0378}{\includegraphics[scale=0.06]{orcid.pdf}\hspace{1mm}Thomas Mikolajick} \\
    \\
	NaMLab gGmbH\\
	01187 Dresden, Germany \\
    \And
	\href{https://orcid.org/0000-0001-8456-2286}{\includegraphics[scale=0.06]{orcid.pdf}\hspace{1mm}Beatriz Noheda} \\
    Zernike Institute for Advanced Materials \& \\ Groningen Cognitive Systems and Materials Center (CogniGron)\\
	University of Groningen\\
	9747 AG Groningen, Netherlands \\
    \And
	\href{https://orcid.org/0000-0003-0479-6897}{\includegraphics[scale=0.06]{orcid.pdf}\hspace{1mm}Erika Covi^*} \\
    \\
	NaMLab gGmbH\\
	01187 Dresden, Germany \\
}
\else
\usepackage{authblk}

\newbox{\orcid}\sbox{\orcid}{\includegraphics[scale=0.06]{orcid.pdf}} 
\author[1,2]{%
	\href{https://orcid.org/0000-0003-0993-5593}{\usebox{\orcid}\hspace{1mm}Luca Fehlings\thanks{\texttt{Corresponding author: l.d.fehlings@rug.nl}}}%
}
\author[1,3]{%
	\href{https://orcid.org/0000-0003-3814-0378}{\usebox{\orcid}\hspace{1mm}Thomas Mikolajick}%
}
\author[2]{%
	\href{https://orcid.org/0000-0001-8456-2286}{\usebox{\orcid}\hspace{1mm}Beatriz Noheda}%
}
\author[1,2]{%
	\href{https://orcid.org/0000-0003-0479-6897}{\usebox{\orcid}\hspace{1mm}Erika Covi}%
}
\affil[1]{NaMLab gGmbH, 01187 Dresden, Germany}
\affil[2]{Zernike Institute for Advanced Materials \& Groningen Cognitive Systems and Materials Center (CogniGron)
	University of Groningen, 9747 AG Groningen, Netherlands}
\affil[3]{Technical University of Dresden, 01187 Dresden, Germany}
\fi

\renewcommand{\headeright}{}
\renewcommand{\undertitle}{}
\renewcommand{\shorttitle}{Millisecond-scale volatile memory in HZO ferroelectric capacitors for bio-inspired temporal computing}

\hypersetup{
pdftitle={Millisecond-scale volatile memory in HZO ferroelectric capacitors for bio-inspired temporal computing},
pdfsubject={cond-mat.mtrl-sci, eess, cs.ET},
pdfauthor={Luca Fehlings},
pdfkeywords={hzo, hafnia, ferroelectrics, retention, time scales, synapse},
}

\usepackage{xcolor}
\usepackage[acronym]{glossaries}
\newacronym{beol}{BEOL}{back-end of the line}
\newacronym{rsnn}{rSNN}{recurrent spiking neural network}
\newacronym{hzo}{HZO}{Hf${_{0.5}}$Zr${_{0.5}}$O${_2}$}
\newacronym{nbox}{NbO$\mathrm{_x}$}{reactively sputtered, oxygen-deficient niobium oxide}
\newacronym{pzt}{PZT}{Pb[Zr${\mathrm{_x}}$Ti$\mathrm{_{1-x}}$]O$_3$}
\newacronym{pund}{PUND}{positive-up-negative-down}
\newacronym{fecap}{FeCap}{ferroelectric capacitor}
\newacronym{cmos}{CMOS}{complementary metal oxide semiconductor}

\begin{document}
\maketitle

\begin{abstract}
        With the broad recent research on ferroelectric hafnium oxide for non-volatile memory technology, depolarization effects in HfO$_2$-based ferroelectric devices gained a lot of interest. Understanding the physical mechanisms regulating the retention of these devices provides an excellent opportunity for device optimization both towards non-volatile memory applications and towards real-time signal processing applications in which controlled time constants are of paramount importance. Indeed, we argue that ferroelectric devices, particularly HfO$_2$-based, are an elegant solution to realize possibly arbitrary time constants in a single scaled memory device, which paves the way for temporal and brain-inspired computing in hardware. Here we present a ferroelectric capacitor stack realizing volatile memory due to its unique interface configuration. We provide electrical characterization of the device to motivate its use for realizing time constants in hardware,
        followed by an investigation of the electronic mechanisms and their possible relation to the observed retention times to facilitate further modeling of the retention process in HfO$_2$-based ferroelectric capacitors. In the presented device, internal electric fields stabilize one polarization of the ferroelectric film, opening the possibility for unipolar operation with millisecond retention for the unstable polarization state. We show a dependence of the retention on both the polarization as well as the electrical stimuli, allowing us to exploit a range of time scales in a single device. Further, the intentionally defective interface in the presented material stack allows an insight into the interplay between retention loss in HfO$_2$-based ferroelectric devices and the internal bias field, which we relate to the interface composition and the role of oxygen vacancies as a possible source of the internal bias fields.
    \end{abstract}
    
    %\begin{IEEEkeywords}
    %    HZO, volatile memory, neuromorphic computing, multi-time scale, retention, depolarization
    %\end{IEEEkeywords}
    
    %----------------------- Introduction -----------------------
    \section{Introduction}

        Memory devices that exhibit limited but controlled, and ideally tunable, memory retention have promising applications in dynamical systems that need specific time constants, such as spiking neural networks or control systems. Brain-inspired computational mechanisms such neural adaptation, eligibility traces, three-factor learning rules and short-term plasticity \cite{susman_stable_2019, mongillo_intrinsic_2017} require analog and temporal encoding of information. In these systems, analog memory elements with varying memory retention \cite{tetzlaff2012} need to be implemented and addressed asynchronously. Software and digital \gls{cmos} implementations typically require a high time resolution, leading to limited scalability due to finite computation power \cite{vanalbada2018}. Analog-mixed-signal \gls{cmos} solutions make use of integrated capacitors and current-mode or subthreshold circuits \cite{Chicca2014}, however, these circuits suffer from scaling and reliability issues hindering large-scale systems. Here, we aim to exploit the underlying physical mechanisms of retention to directly observe the retention loss transient within experimentally accessible timescales, thereby enabling more accurate modeling of both retention loss and depolarization dynamics. The effective retention of charge-based memory devices, and ferroelectric devices in particular, depends on the material stack and defect-related mechanisms such as charge trapping, dipoles and ionization effects within these materials. These mechanisms can then be exploited in two ways: One is to improve the retention characteristics towards the 10-year retention goal for non-volatile memory applications. The other, still largely unexplored in devices based on ferroelectric HfO$_2$, is to control the retention and purposefully engineer the device volatility in time scales that are relevant for applications such as hardware implementations of synaptic plasticity or neuronal membrane potentials. We argue that we can relate the time constant of the device directly to the physical mechanism of the ferroelectric polarization process, leading to a hardware time constant that is, for example, independent of device area. 
        
        Ferroelectric hafnium oxide thin films are a promising material system for dense non-volatile memory arrays, due to their scalability, fast switching times, back-end-of-line (BEOL) \cite{francois2019} and front-end-of-line (FEOL) \cite{dünkel2017} \gls{cmos} process compatibility, and power efficiency \cite{miko2020}. Additionally, their unique switching dynamics and polycrystalline nature allow ferroelectric hafnium oxide to be used as an analog memory element with gradual switching characteristics \cite{gong_nucleation_2018}. Unique wake-up behavior based on internal bias fields \cite{schenk_complex_2015}, phase transformations \cite{grimley_tetr_interface} and imprint mechanisms \cite{florent2017reliability} are observed in this material, attributed to the multi-phase material system and charge trapping at defect sites. Moreover, one of the key performance indicators for the non-volatile memory market is data retention for more than 10 years under high temperature storage conditions. In HfO$_2$-based ferroelectric devices, due to the thickness of the ferroelectric typically being in the range of 10\,nm, this is challenging due to depolarization effects commonly originating from non-ideal charge screening \cite{zhao_depolarization_2019} and imprint effects \cite{vishnumurthy2024ferroelectric}. The non-ideal charge screening is critical for HfO$_2$-based ferroelectrics, as parasitic interface capacitances due to a paraelectric tetragonal HfO$_2$-based phase \cite{grimley_tetr_interface}, oxidized electrodes \cite{lomenzo_2015}, or finite screening in the metal or semiconducting electrodes \cite{black_1999} can be on the same order of magnitude in film thickness and capacitance as the ferroelectric layer itself, leading to a depolarization field and a trade-off in operating voltage and retention when designing the material stack.  In addition, the imprint mechanism leads to higher required voltages for reading out the polarization and to a limited retention of the state opposite to the programmed state during imprinting, commonly referred to as opposite state retention \cite{alcala_2022, mueller_2013}.
        
        Here, we introduce an additional 15\,nm thick, reactively sputtered \gls{nbox} layer into a conventional \gls{hzo}-\gls{fecap} stack. Thin films of \gls{nbox} show several unique features, particularly when doped or defective, such as trap-based conduction \cite{jouv_electrical_1991} and high mobility of oxygen ions \cite{fuschillo_dielectric_1975}. Indeed for reactively sputtered and amorphous thin films, the ratio of the Nb$_2$O$_5$ phase fraction, and hence conductance and polarizability, can be tuned continuously via the oxygen partial pressure during deposition \cite{graca_electrical_2015}. For ferroelectric devices, ALD-Nb$_2$O$_5$ has been used previously in devices as an interlayer to enhance the remanent polarization and reduce wake-up effects \cite{m_i_popovici_high_2022} in HfO$_2$-based films, attributed to its ability to oxidize the oxygen vacancies at the HfO$_2$ interface and therefore reduce the amount of defects in the ferroelectric layer. Further studies have shown that sputtered Nb$_2$O$_5$ can increase the oxygen vacancy density in the layers depending on which metal or oxide layer it interfaces \cite{engl2025}. Thin films of \gls{nbox} have also been utilized as an active layer for memristive devices, where, besides realizing the well known valence change mechanism \cite{MAHNE201273}, they can be used to realize threshold switching devices \cite{ascoli2021local}, analog switching where oxygen vacancies accumulate at an Al$_2$O$_3$ interface \cite{mgeladze_2022} or act as field-effect passivation layers with a high fixed charge density at the interface to SiO$_2$ \cite{macco_effective_2018}. Here we exploit this effect by introducing the Al/\gls{nbox} layers, whose unique interaction of the oxygen-deficient NbOx with the \gls{hzo} layer generates oxygen vacancies at the top electrode interface, leading to a strong asymmetry in the density of charged defects between the top and bottom electrode. This defect configuration leads to strong and directed internal bias fields already observed for thinner \gls{nbox} layers\cite{engl2025}, which, in this work, we relate to the measured retention as well as the asymmetry in the hysteresis curve and the field cycling behavior.
        
        In our work, we explore the potential of controlled volatility in HfO$_2$-based ferroelectric devices and relate them to the physical mechanisms regulating retention in HfO$_2$ ferroelectrics. We introduce an additional \gls{nbox} layer with an Al top electrode to a ferroelectric capacitor stack, leading to an asymmetric polarization hysteresis and a resulting retention in the time window of 0.1\,ms to 2\,ms. Millisecond time-scales have already been realized on a device level in filamentary volatile devices, with tunability of the retention via current compliance in 1T-1R cell \cite{e_covi_switching_2021}. However, due to their switching process, relying on the surface tension of the mobile ion species forming the filament, the process is inherently stochastic, and thus the distribution of the retention time spans over several orders of magnitude \cite{wang_switching_2021}. Compared to state-of-the-art solutions based on resistive switching devices \cite{Ricci_2023}, we aim to employ a less stochastic, although not as tunable, device based on the switching dynamics of ferroelectric capacitors.

    %----------------------- Methods -----------------------
    \section{Methods}
        %----- Figure 1 -----
        \begin{figure}[t]
            \centering
            \includegraphics[]{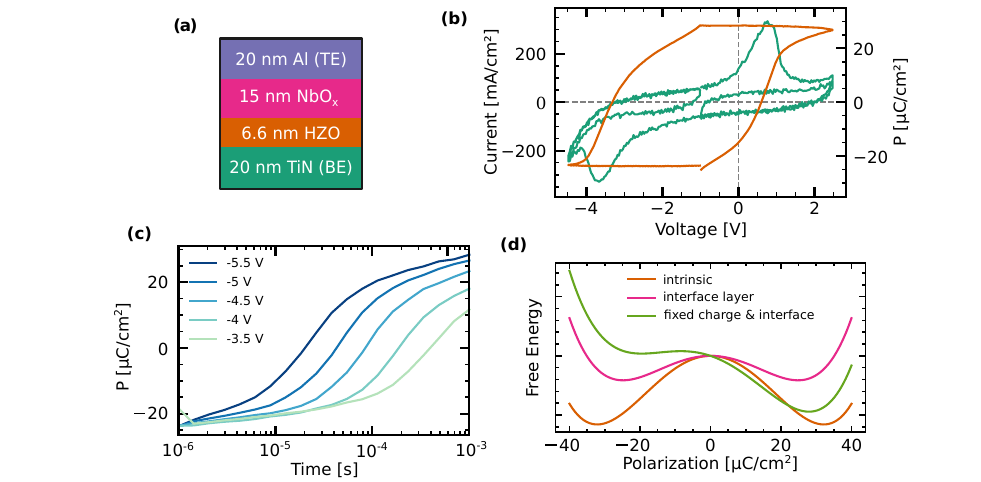}
            \caption{(a) Material stack of the device. (b) Continuous positive-up-negative-down (PUND) measurement at 1\,kHz of the device after 1000 wake-up cycles, leading to a strongly asymmetric transient I-V characteristic and corresponding polarization hysteresis loop. The P$_\uparrow$ state is unstable as it reverses at 0\,V bias (dotted line). (c) Switching kinetics of the device towards the P$_\uparrow$ state, showing switching over several orders of magnitude in time. (d) Illustration of the effect of interface layers and an internal bias field arising from fixed charges on the free energy landscape of a ferroelectric, based on Eq. S(4). For both an interface layer and fixed charges, the ferroelectric can become effectively unipolar with one polarization state depolarizing depending on the energy barrier.} 
            \label{fig:1}
        \end{figure}
        
        The device consists of a ferroelectric Hf${_{0.5}}$Zr${_{0.5}}$O${_2}$ (\gls{hzo}) layer with asymmetric electrodes, as sketched in Fig.~\ref{fig:1} (a). The \gls{hzo} layer of 6.6\,nm thickness was deposited via ALD at 280\,$^{\circ}$C using alternating cycles of HyALD and ZyALD precursors with O$_3$ as reactive gas. The bottom electrodes are formed from TiN with a thickness of 20\,nm each. While at the bottom electrode the TiN is in direct contact to the \gls{hzo}, at the top electrode a 15\,nm \gls{nbox} layer deposited by reactive sputtering is placed between \gls{hzo} and an aluminum electrode. The \gls{nbox} was sputtered at \mbox{5e-3\,mbar} pressure with 45\,sccm Ar and 12\,sccm O$_2$ flow from a Nb target. As the sputtered \gls{nbox}, in contrast to bulk Nb$_2$O$_5$, is defective and oxygen-deficient it is doped and therefore conductive \cite{jouv_electrical_1991, fuschillo_dielectric_1975}. The whole stack was finally annealed at 500\,$^{\circ}$C for 20\,s in a N$_2$ atmosphere to crystallize the \gls{hzo} layer. 
        Both top and bottom electrodes are patterned via UV-lithography, resulting in devices with different active areas ranging from 25\,\textmu m\,$\times$\,25\,\textmu m down to  5\,\textmu m\,$\times$\,5\,\textmu m. Unless otherwise noted, in the following we show the results obtained measuring devices with an active area of 25\,\textmu m\,$\times$\,25\,\textmu m.
        
        Electrical characterization was carried out using a Keithley 4200A semiconductor parameter analyzer with 4225-PMU units for the transient measurements. All measurements were carried out under ambient conditions, and voltages refer from top to bottom electrode, where Al/NbOx is referred to as the top electrode (Fig. \ref{fig:1} (a)).

    \section{Results}
    \subsection{Electrical characterization}

        %----- Figure 2 -----
        \begin{figure}[t]
            \centering
            \includegraphics[]{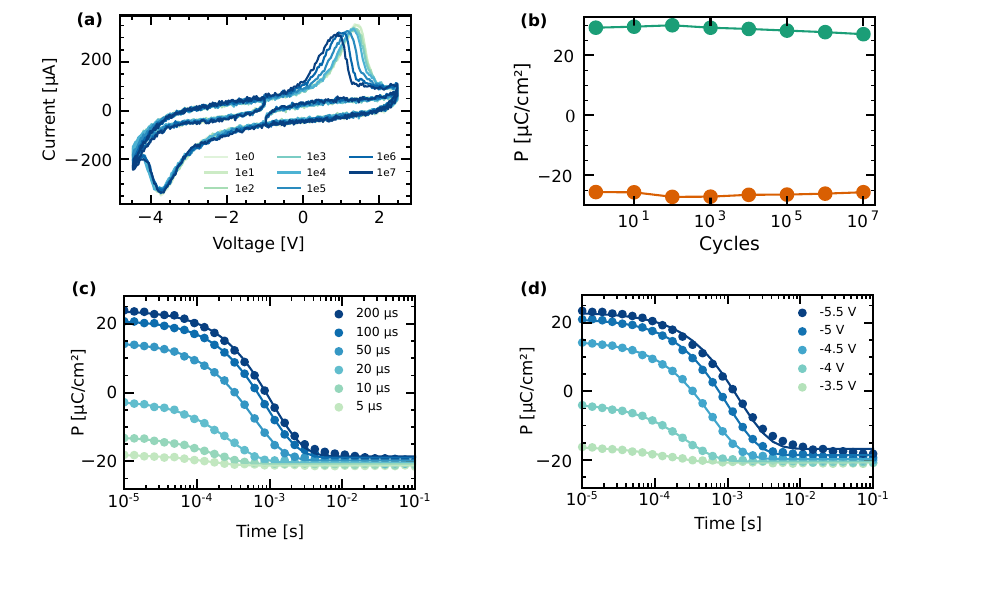}
            \caption{(a) Transient I-V measurement for endurance cycling at a 100 kHz triangular wave showing no significant fatigue over $10^7$ cycles. Notably the current peak at positive voltages migrates to lower voltages. (b) Remanent polarization over cycles for $10^7$ cycles at 100 kHz cycling between -4.5\,V and 2.5\,V. (c) Measured retention (dots) of the  P$_\uparrow$ state after programming with a pulse of -4.5\,V amplitude and different pulse widths. For wider pulses, both the programmed polarization as well as the retention increases. The lines show respective exponential fits following Eq.~\ref{eq:exponential}. (d) Retention for programming pulses with a fixed width of 50\,\textmu s and increasing amplitudes, showing results similar to the pulse width variation, with a notable increase in the tail value for longer times.}
            \label{fig:2}
        \end{figure}

        As evident from the \gls{pund}-measurement (Fig.~\ref{fig:1} (b)), the woken-up device has a highly asymmetric polarization hysteresis, with the ferroelectric polarization already switching to P$_\downarrow$ at 0\,V. In addition, there is a significant hysteresis between the current for the switching cycle and the non-switching cycle, apart from the switching peak itself. This implies either a strong broadening of the switching peaks or an additional hysteretic process, such as charge trapping. For clarity, the \gls{pund} measurements here are performed symmetrically around -1\,V, as with a conventional measurement symmetric around 0\,V, the polarization would partly reverse during the sweep back. 

        Based on the switching kinetics (Fig.~\ref{fig:1}(c)), it is also evident that the device can be switched gradually using different combinations of write pulse amplitudes and time widths, as expected for micrometer-sized \gls{hzo} \glspl{fecap}. The switching kinetics however are asymmetric between polarities and the switching in significantly slower for negative voltages due to the asymmetry of the polarization hysteresis. Fig. \ref{fig:1} (d) illustrates this asymmetry based on the Landau free energy model: Due to the \gls{nbox} interface layer and positive fixed charges within the \gls{nbox} layer or its interface with the \gls{hzo} layer, the energy landscape becomes skewed towards one polarization. The energy barrier to transition from that polarization, in this case P$_\uparrow$ to P$_\downarrow$, is then low enough to allow a thermal transition with no externally applied field. With further endurance cycling (Fig.~\ref{fig:2}(a)), the positive voltage peak in particular drifts to lower voltages. There is, however, no significant fatigue in the total amount of polarization up to 10$^7$ cycles at 10\,kHz between -4.5\,V and 2.5\,V (see Fig.~\ref{fig:2}(b)). It should be noted that for higher voltages, ranges, and high cycle counts there is evident fatigue and an increase in leakage current, in addition to peak splitting and shifting (Fig.~S3-S5).

    %----------------------- Retention -----------------------
    \subsection{Retention characteristics}

        Ferroelectric memory devices based on polycrystalline doped HfO$_2$ thin films show unique switching behavior attributed to the dominant role of domain nucleation in the switching process\cite{gong_nucleation_2018}, distinctive from perovskite ferroelectrics like \gls{pzt}. Due to their low thickness, typically between 6\,nm and 12\,nm, and their ability to gradually switch even at low electric fields, \gls{hzo}-based ferroelectric memory devices are severely impacted by internal electric fields, arising from charge trapping and non-ideal charge screening at the interface with the electrodes \cite{chouprik_origin_2021, zhao_depolarization_2019}. Apart from the polarization switching process itself, the polarization retention is particularly affected by these depolarization fields, which commonly arise from non-ideal interfaces intrinsic to the fabrication processes used, for example paraelectric non-switching dead layers in series with the ferroelectric phase \cite{grimley_tetr_interface} or are intrinsic to material stack, such as a limited screening length in the metal electrodes \cite{black_1999}.
        To model the retention of a ferroelectric capacitor, there are two major mechanisms (Eq. \ref{eq:e_t}) to consider: an electric field $E_{dep}$ proportional to the polarization $P$ that arises due to the imperfect charge screening at the interface, and an internal bias $E_{bias}$ arising from several charge-based effects, such as ionized defects, trapped charges, or a work function difference between the contacting electrodes \cite{schenk_complex_2015}
        
        \begin{equation}
            E(t) = E_{bias} + E_{dep}(t)\,.
            \label{eq:e_t}
        \end{equation}
        The depolarization field is anti-parallel to the field applied to switch the \gls{fecap}, leading to a negative feedback:
        \begin{equation}
            E_{dep}(t) = -\frac{P(t)}{\epsilon_0 \epsilon_{FE}} \cdot \frac{1}{1+(C_{s}/C_{FE})}\,,
            \label{eq:e_dep}
        \end{equation}
        
        where $\epsilon_{FE}$ and $C_{FE}$ are the relative permittivity and capacitance of the ferroelectric layer, $C_s$ is the capacitance of the series capacitor, and $P(t)$ is the ferroelectric polarization, bound by the saturation polarization $\pm P_s$. Accordingly, the retention of the device is dictated by the charge distribution inside the material stack and is therefore independent of the device area, in contrast to the area-dependent RC time constant of a conventional capacitor. 
        
        For the volatile \gls{fecap}, the retention is investigated based on the programming conditions (Fig.~\ref{fig:2}(c),(d)) and generally shows polarization loss at no applied bias. In each case, the device is pre-polarized to the negative saturation polarization and then polarized using the indicated pulse width and amplitude, where both applied pulse width (Fig.~\ref{fig:2}(c)) and amplitude (Fig.~\ref{fig:2}(d)) are varied. Then, the polarization loss is measured, showing a retention in the range of 0.1\,ms to 2\,ms, increasing with pulse width, amplitude and programmed initial polarization. In contrast to conventional depolarization based on $E_{dep}$, the device does not depolarize to the  electrostatically neutral polarization, but rather fully reverses to the opposite saturated polarization state. This stable polarization state in turn shows no observable depolarization behavior (Fig.~S6). Under a field $E_{dep}$ due to imperfect charge screening at the interfaces, as described in Eq.~\ref{eq:e_dep}, the polarization can only depolarize to P\,=\,0, as the depolarization field is always of opposite polarity of the ferroelectric polarization. Hence, the retention behavior observed here has to result partly from a static bias field $E_{bias}$ that effectively stabilizes the negative saturation polarization at no applied field and reverses the positive polarization state as observed in the retention measurements. 
        
\begin{figure}[t]
            \centering
            \includegraphics[]{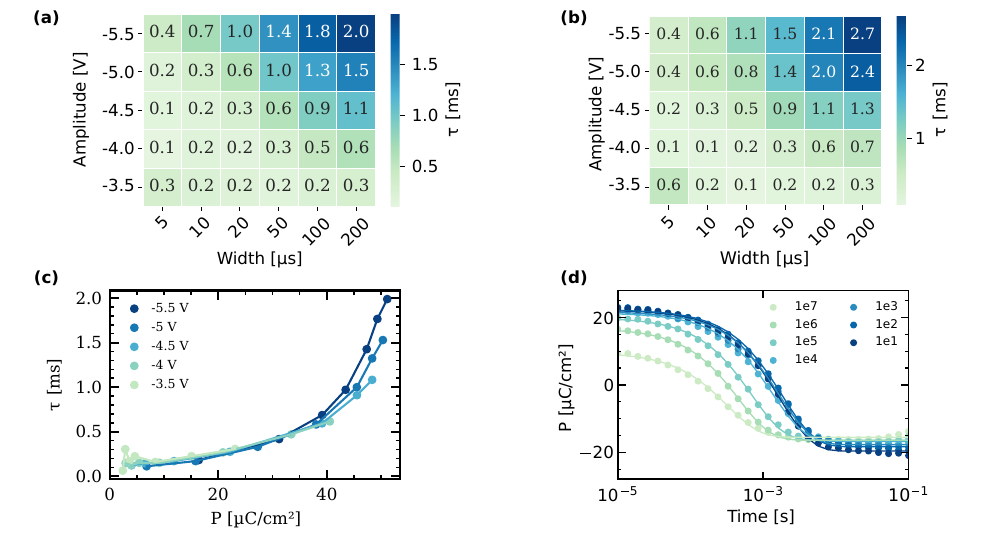}
            \caption{(a) Time constant as a function of the width and amplitude of the programming pulse. Both pulse width and amplitude can be utilized to achieve time constants between 0.1\,ms and 2\,ms in the explored parameter range. (b) Repetition of the previous retention for a 5\, \textmu m $\times$ 5\, \textmu m area device, i.e. 25 times smaller area device, showing similar retention times and reinforcing claim of an area-independent time constant (c) The time constant is not purely a function of polarization. For voltages above -4.5\,V, the correlation between polarization and time constant ceases and the time constant begins to increase rapidly due to the modulation of the internal bias field by the combination of high pulse amplitude and width. (d) Retention of the P$_\uparrow$ state over endurance cycling showing a decrease in initial polarization while the steady-state polarization increases.}
            \label{fig:3}
        \end{figure}

        The time constant of the retention can be defined analogous to a dielectric capacitor via an exponential decay of the polarization:
        
        \begin{equation}
            P(t) = P_0 \cdot  \mathrm{exp} \left(-\frac{t}{\tau}\right) + P_{\infty}\,,
            \label{eq:exponential}
        \end{equation}
        
        where $P_{\infty}$ is the steady state polarization, the sum of $P_{\infty}$ and $P_0$ is the initial polarization, and $\tau$ is the time constant. As evident by Fig.~\ref{fig:2} (c), the exponential decay fits the retention behavior, with exception of the late-phase retention for longer pulse width, which will be considered later. For the selected range of programming pulse widths and amplitudes, the resulting time constants (Fig.~\ref{fig:3} (a)) reveal a general trend of increasing time constants with both pulse width and amplitude. In fact, the same retention behavior at the same time constant has been observed for smaller devices (Fig.\ref{fig:3} (b)), indicating that the time constant is independent of device size, at least until different scaling effects set in, e.g., if the device size approaches the grain size. 
        
        As not all amplitude/width permutations fully polarize the device, a correlation of the time constant on the initial polarization $P_{\infty} + P_0$ is close at hand and is indeed partially confirmed as seen in Fig.~\ref{fig:3} (c). This retention penalty for unsaturated polarization states has already been observed for conventional HfO$_2$-based \glspl{fecap} \cite{ chouprik_origin_2021, muller_ferroelectric_2013}, and a common physical origin could be further elucidated by our findings. However, for voltages \mbox{-5\,V} and \mbox{-5.5\,V}, an additional increase in the time constant can be seen in the regime of the fully saturated polarization. Recalling the only partial polarization reversal for high voltages shown in Fig.~\ref{fig:2} (d), a common explanation could be a modulation of the internal bias by the high-amplitude pulses, so that the equilibrium polarization is shifted slightly above the negative saturation polarization. It should be noted that this effect is reversible, i.e., after a reset operation, the original retention behavior is restored (Fig.~S1,S2).

        %----- Figure 3 -----

    %----------------------- Discussion -----------------------
    \section{Discussion}

        %----- Figure 4 -----
        \begin{figure}[t]
            \centering
            \includegraphics[]{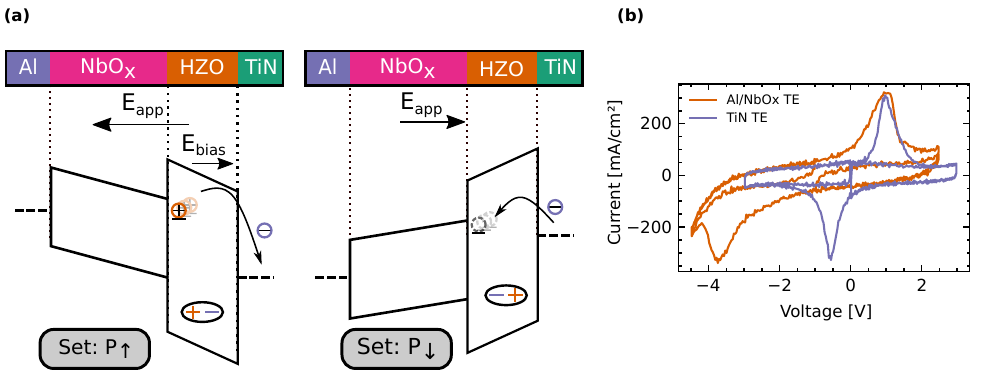}
            \caption{(a) Band diagrams of the switching process into the P$_\uparrow$ and P$_\downarrow$ states. While the switching process for the P$_\downarrow$ state is close to that of a conventional ferroelectric capacitor, switching into the P$_\uparrow$ process involves as de-trapping of electrons, which happens in parallel with the switching process and creates an opposing bias field. This leads to the broad switching peak observed. (b) Comparioson of two devices with a convention TiN electrode and the investigated Al/NbOx electrode. In comparison, the device with the Al/NbOx electrode shows broader switching peaks due to the trapping process. The negative switching peak is broadened due to the positive charges created in parallel with the switching process, increasing the opposing bias field. }
            \label{fig:4}
        \end{figure}
    
        The investigated \gls{fecap} shows a unique asymmetric polarization with near-absent wake-up as well as a millisecond retention for the P$_\uparrow$ state, with no observed retention loss for the P$_\downarrow$ state. As this is a material stack to intentionally induce volatility in a \gls{fecap}, a deeper look into the peculiarities of this material stack is needed. In comparison to a conventional TiN/HZO/TiN \gls{fecap} stack (Fig.~\ref{fig:4}(b)), the investigated Al/\gls{nbox} configuration poses some fundamental differences. 
        
        Aluminum as an electrode material has a low work function and is known to react with oxygen-containing materials by forming a thin Al$_2$O$_3$ interface layer. \gls{nbox} thin films on the other hand are known for their high concentration of meta-stable oxygen vacancies and the high mobility of oxygen ions in the thin film \cite{fuschillo_dielectric_1975}. In the context of HZO \glspl{fecap}, oxygen vacancies have already proven to be central to the formation of the ferroelectric orthorhombic phase, as well as the characteristic wake-up behavior attributed to domain (de-)pinning and t $\rightarrow$ o phase transformations \cite{schenk_complex_2015, chouprik_wake-up_2019}. Bi-layer Nb$_2$O$_5$-La:HZO stacks based on ALD-deposited Nb$_2$O$_5$ have already been demonstrated to have improved wake-up and higher remanent polarization. This was attributed to migration of oxygen vacancies from HZO to Nb$_2$O$_5$ and a corresponding decrease in the t-phase fraction at the interface \cite{m_i_popovici_high_2022}. 

        In our case, the polarization hysteresis and retention measurements suggest an internal positive bias field opposing the P$_\uparrow$ polarity, while the P$_\downarrow$ is unaffected, as can be seen in comparison with a device with a TiN electrode (Fig.~\ref{fig:4}(b)). Accordingly, assuming an increased oxygen vacancy density at the \gls{nbox} interface, a mechanistic explanation of the internal bias could be due to the excitation\,/\,relaxation of the trapping sites n from the oxygen vacancies \cite{foster2002vacancy}. These trapping mechanisms can lead to pinning of domains and associated internal bias fields \cite{schenk_complex_2015} and have been identified as key mechanisms in the wake-up and anti-ferroelectric behavior of HfO$_2$-based \glspl{fecap} \cite{chouprik_wake-up_2019}, which highlights the importance of the interplay of internal bias fields and oxygen vacancies. Here, the mechanism is illustrated in Fig.~\ref{fig:4} (a): When applying a negative voltage (set P$_\uparrow$) the domains switch, as indicated by the dipole, and the electrons on the \gls{nbox} interface de-trap, leaving behind positively charged oxygen vacancies. As soon as the applied field is removed, the internal fields are dictated by the ferroelectric dipoles and the charged defects. Assuming that the ferroelectric dipole is screened by charges at the interface, the positively charged oxygen vacancies create an electric field that reverses the polarization. As soon as a positive voltage is applied (set P$_\downarrow$), the domains switch and electrons trap at the \gls{nbox} interface and thus electrically neutralize the defects. When the electric field is removed, again the defect charges and the depolarization field build up an internal bias. In this case however, the bias field is not sufficient to reverse the polarization. This could be either due to significantly more oxygen vacancies being present at the \gls{nbox} interface or to an easier electron trapping process to neutralize them. As the \gls{nbox} film is already oxygen-deficient after deposition, a high oxygen vacancy concentration is plausible, particularly considering the annealing process that is performed on the full Al/\gls{nbox}/\gls{hzo}/TiN stack.
        
        The behavior elaborated here can also be related to short term imprint effect in HfO$_2$-based \glspl{fecap} arising from internal bias fields: While P$_\downarrow$ easily imprints and therefore stabilizes as seen in conventional FeCaps, P$_\uparrow$ is not able to stabilize due to both the high defect density at the \gls{nbox} interface and possibly the unlikely electron tunneling due to the \gls{nbox} interface in the short time-frame, i.e., before the reversal of the ferroelectric domains by the internal bias field. Additionally, the missing wake-up and no visible pinch in the hysteresis loop (Fig.~\ref{fig:2}(a)) could be explained by the oxygen vacancy distribution in Fig.~\ref{fig:4}(a). Since the charged defects are highly asymmetric in our stack, the opposing internal bias fields potentially causing the pinching via static or pinned domains could be skewed in one direction, in contrast to the device with a TiN top electrode where bias fields in both directions can occur. Further, the shift of the equilibrium polarization and the modulation of the retention time by high set pulse amplitudes and widths could be explained by a change in oxygen vacancy concentration or trapping rate. This could lead to some ferroelectric domains having a significantly reduced internal bias field due to less positive charges originating from oxygen vacancies.
        
        In addition to the retention mechanism, the increased leakage current compared to the conventional TiN electrode (Fig.~\ref{fig:4}) implies a significant change in the material composition. This can be supported by the previously mentioned creation of defects in the \gls{hzo} via the \gls{nbox} layer. The creation of oxygen vacancies could aid a trap-based conduction mechanism, causing higher leakage currents. This can also explain the higher observed remanent polarization in the device with the \gls{nbox} interlayer compared to the convention device, as the polar orthorhombic phase is associated with oxygen vacancies. In particular for the 6.6\,nm thick \gls{hzo} layer in this device, the additional oxygen vacancies could assist in stabilizing the orthorhombic phase in this layer which is thinner than the most frequently used 10\,nm \gls{hzo} layer.

        Further analysis on this control of the bias field by NbOx-induced oxygen vacancies could enable \glspl{fecap} with time constants tuned by the material properties or stack design of the devices. Longer time constants could be achieved my increasing the \gls{hzo} thickness, with the caveat of increasing the operation voltage, or by tuning the oxygen content in the \gls{nbox} layer, either by the deposition process directly or by exploration of electrode materials other than Al. This could enable sophisticated control of the bias field, and thus retention, of the device. As conventional FeCaps have shown retention above 10 years, any retention time could be achieved if the oxygen vacancy concentration can be controlled reliably. Accordingly, time constants higher or lower than the ones present here could be achieved, allowing exploration of several applications where these bio-inspired time constants \cite{tetzlaff2012} could be utilized. Due to the wake-up-free operation and potential scalability of the presented device, these devices could be applied in hardware realizations of multi-timescale algorithms such as working memory \cite{Ricci_2023} or two-phase synaptic plasticity \cite{luboeinski2023organization, atoui2025multi}.

    %----------------------- Application -----------------------
    
    \section{Conclusion}
        By introducing an Al/\gls{nbox} top electrode to a \gls{hzo}-based \gls{fecap}, we observe a strong asymmetry in the polarization hysteresis, notably by a strong shift of the negative switching peak towards higher absolute voltages. This asymmetry manifests in its volatile polarization in the order of milliseconds, where the retention time is a function of both the total polarization, i.e., the amount of switched domains, and the amplitude of the applied set pulse. As the device fully reverses from the unstable P$_\uparrow$ to the stable P$_\downarrow$ polarization, we identify an internal bias field as the root cause of the asymmetry and the volatile behavior observed. This internal bias field is attributed to a high concentration of oxygen vacancies at the \gls{nbox}/\gls{hzo} interface, resulting from the oxygen-deficient \gls{nbox} layer. These oxygen vacancies, when relaxed, act as positive charges from which the bias field originates, in accordance with previous investigations on wake-up and imprint. This electronic mechanism also explains other characteristics of the observed device, such as the missing wake-up, the unipolar peak shifts, and the increased leakage current that is non-linear with area.
        
        In addition to the further insight offered on the relation of internal bias field to oxygen vacancies in \gls{hzo}-based \gls{fecap}, the concept of engineering the retention of a \gls{fecap} using the proposed materials and mechanisms provide a promising route to achieve medium time constants in the range of milliseconds to minutes, between capacitors and conventional non-volatile memories. With the good integrability of \gls{hzo} with \gls{cmos} technology and the nearly absent wake-up, this device is a good candidate to be integrated into circuits for which biological time constants or medium time constant dynamics are required, such as bio-inspired in neural networks or control systems.

    %\section*{Data availability}
    %\section*{Code availability}
    \section*{Acknowledgments}
    This work was supported by the European Research Council (ERC) through the European’s Union Horizon Europe Research and Innovation Programme under Grant Agreement No 101042585. Views and opinions expressed are however those of the authors only and do not necessarily reflect those of the European Union or the European Research Council. Neither the European Union nor the granting authority can be held responsible for them. The University of Groningen would like to acknowledge the financial support of the CogniGron research center and the Ubbo Emmius Fund.
    %\section*{Author contributions}
    %\section*{Competing Interests}

\bibliographystyle{ieeetr}
\bibliography{refs}

\setcounter{figure}{0}
\renewcommand{\figurename}{Fig.}
\renewcommand{\thefigure}{S\arabic{figure}}
\include{supplementary}

\end{document}

%% file: supplementary.tex
\newpage
\section*{Supplementary Information}

\subsection*{Equations used in Fig.~\ref{fig:1}(d)}
    Landau-Devonshire with depolarization:
    \begin{equation}
        \frac{F_i}{V_*} = \frac{\alpha}{2} \cdot D_i^2 + \frac{\beta}{4} \cdot D_i^4 + \gamma \cdot D_i^2 - D_i \cdot E_{bias} \cdot \mathrm{cos}(\theta)
        \label{eq:landau_devonshire}
    \end{equation}
    
    Depolarization factor:
    \begin{equation}
        \gamma = -\frac{d_{int}}{d_{FE} \cdot \epsilon_0 \cdot \epsilon_{int}}
        \label{eq:ld_dep_factor}
    \end{equation}

    \begin{table}[h]
        \centering
        \caption{Landau-Devonshire parameters used in Eqs. (4) and (5)}
        \begin{tabular}{lllll}
        \toprule
        Parameter & intrinsic & interface & fixed charge + interface \\
        \midrule
        $\mathrm{\alpha}$ & -2.242e8 $\mathrm{Jm/C^2}$ & -2.242e8 $\mathrm{Jm/C^2}$ & -2.242e8 $\mathrm{Jm/C^2}$ \\
        $\mathrm{\beta}$ & 2.170e9 $\mathrm{Jm^5/C^4}$ & 2.170e9 $\mathrm{Jm^5/C^4}$ & 2.170e9 $\mathrm{Jm^5/C^4}$ \\
        $\mathrm{E_{bias}}$ & 0 & 0 & 100 ksV/cm \\
        $\mathrm{d_{int}}$ & 0.2\,nm & 0 & 0.2\,nm \\
        $\mathrm{d_{fe}}$ & 6.6\,nm & 6.6\,nm  & 6.6\,nm  \\
        $\mathrm{\epsilon_{int}}$ & 75 & 0 & 75 \\
        $\mathrm{\theta}$ & 0 & 0 & 0 \\
        \bottomrule
        \end{tabular}
        \label{tab:methods:devices}
    \end{table}

\subsection*{Supplementary figures}

    \begin{figure}[h]
        \centering
        \label{fig:supp_retention_volt_re}
        \includegraphics[width=2.5in]{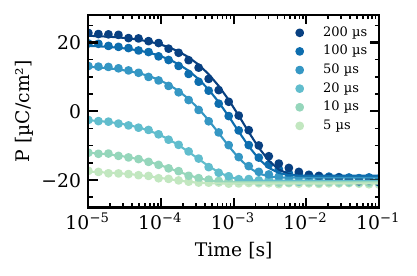}
        \caption{Repetition of measurements in Fig. \ref{fig:2} (c) on the same device, showing that the modulation of the time constant and of the equilibrium polarization after high voltage programming is reversible.}
    \end{figure}
    
    \begin{figure}[h]
        \centering
        \label{fig:supp_retention_time_re}
        \includegraphics[width=2.5in]{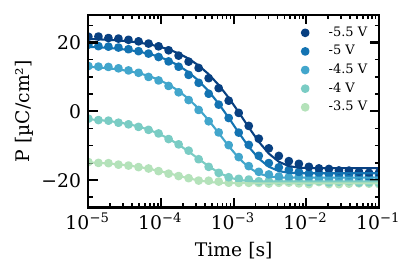}    
        \caption{Repetition of measurements in Fig. \ref{fig:2} (d) on the same device, showing that the modulation of the time constant and of the equilibrium polarization after high voltage programming is reversible.}
    \end{figure}

    \begin{figure}[h]
        \centering
        \label{fig:supp_endurance_high1}
        \includegraphics[width=2.5in]{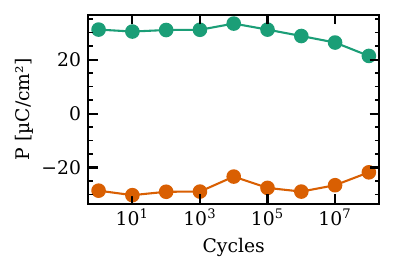}
        \caption{Repetition of the endurance measurement in Fig.~\ref{fig:2}(b) for higher voltages (-5\,V, 3\,V) showing fatiguing behavior.}
    \end{figure}
    
    \begin{figure}[h]
        \centering
        \label{fig:supp_endurance_high2}
        \includegraphics[width=2.5in]{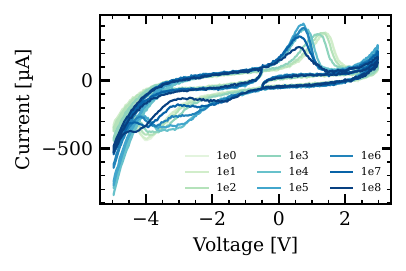}
        \caption{Raw transient I-V data of Fig.~S\ref{fig:supp_endurance_high1}. The leakage current is noticeably higher than in Fig.~\ref{fig:2}(a), although it does not monotonically increase with cycles.}
    \end{figure}
    
    \begin{figure}[h]
        \centering
        \label{fig:supp_endurance2}
        \includegraphics[width=2.5in]{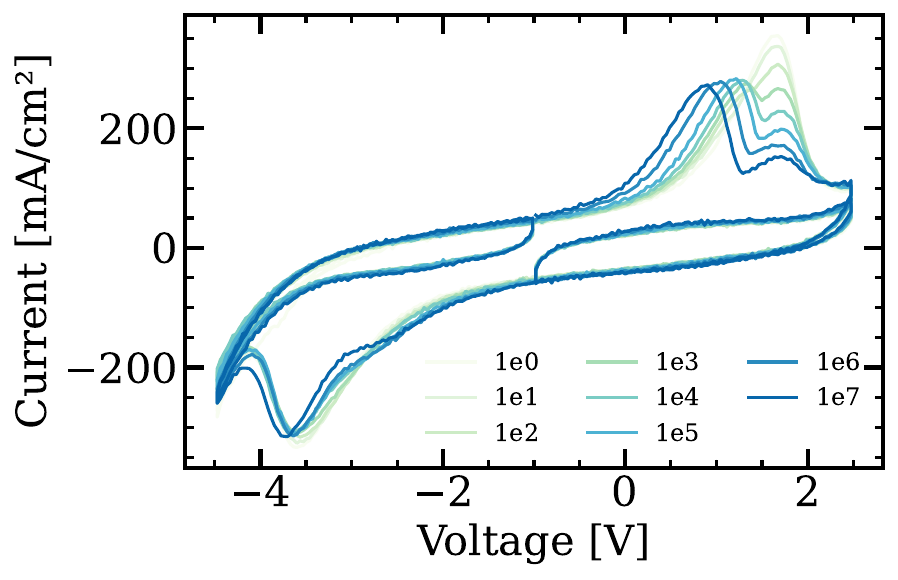}
        \caption{Repetition of the endurance measurement in Fig.~\ref{fig:2}(b) but without a millisecond pause between cycling and PUND measurement. Compared to Fig.~\ref{fig:2}(b), it shows peak splitting, likely as no trap relaxation processes can occur between cycling and PUND measurement.}
    \end{figure}
    
    \begin{figure}[h]
        \centering
        \label{fig:supp_retention_up}
        \includegraphics[width=2.5in]{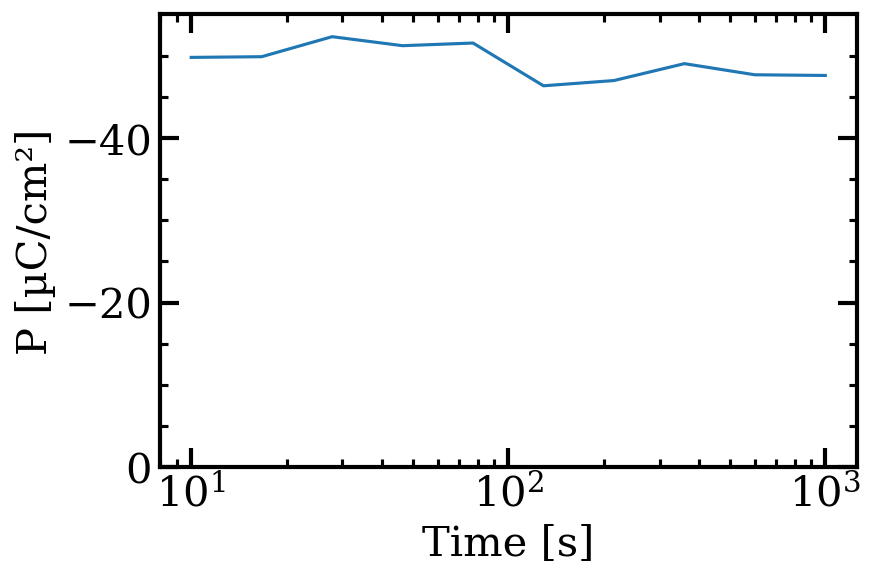}
        \caption{Retention for the stable P$_\downarrow$ state.}
    \end{figure}